\pdfoutput=1
\documentclass[sigplan,nonacm]{acmart}

\setcopyright{none}
\acmConference[ASPLOS '27]{The ACM International Conference on Architectural Support for Programming Languages and Operating Systems}{April 11--15, 2027}{Crete, Greece}

\usepackage{amsmath}
\usepackage{graphicx}
\usepackage{booktabs}
\usepackage{array}
\usepackage{multirow}
\usepackage{tabularx}
\usepackage{tikz}
\usetikzlibrary{arrows.meta,positioning}

\graphicspath{{figures/}}
\newcommand{\sys}{NEURON-Fabric}
\newcommand{\fp}{FP32}
\newcommand{\gbin}{G-Binary}
\newcommand{\gter}{G-Ternary}
\newcolumntype{Y}{>{\raggedright\arraybackslash}X}
\newcommand{\tablebodysetup}{%
  \footnotesize
  \setlength{\tabcolsep}{3pt}%
}

\begin{document}

\title{\sys{}: CXL-Side Low-Bit Gradient Aggregation for Distributed Training}

\author{Ziqiang Wang}
\affiliation{%
  \institution{Carleton University}
  \city{Ottawa}
  \state{Ontario}
  \country{Canada}}
\email{ziqiangwang@cmail.carleton.ca}

\author{Changcheng Huang}
\affiliation{%
  \institution{Carleton University}
  \city{Ottawa}
  \state{Ontario}
  \country{Canada}}
\email{ChangChengHuang@cunet.carleton.ca}

\author{Chung-Horng Lung}
\affiliation{%
  \institution{Carleton University}
  \city{Ottawa}
  \state{Ontario}
  \country{Canada}}
\email{chlung@sce.carleton.ca}

\begin{abstract}
In large-model distributed training, especially large language model (LLM)
workloads, gradient All-Reduce increasingly stresses the memory and
communication path.  This paper asks whether a Compute Express Link (CXL)
memory controller can aggregate low-bit gradient signals as gradient cache
lines pass through it, while preserving a 32-bit floating-point (\fp{}) path for
workloads, layers, or phases that should not use low-bit approximation. We
denote G as the gradient: \gbin{} refers to binary sign-count aggregation of gradient
payloads, while \gter{} indicates ternary-gated gradient aggregation. We present \sys{},
a CXL-side controller architecture that performs these packed aggregation
operations near CXL memory and exposes a simple control interface for selecting
the low-bit or \fp{} path.  Cycle-level timing experiments isolate the
controller datapath and show that the measured five-cycle low-bit aggregation
datapath adds at
most 1.67\% exposed runtime overhead in the full last-level cache (LLC) miss
regime; under bandwidth pressure, the same compute stage is hidden by CXL
service time.  Functional
tests confirm byte-exact identity read-back, \gbin{} sign-count aggregation, and
\gter{} ternary gating.  Separately,
end-to-end training checks quantify the communication and accuracy tradeoff:
low-bit aggregation remains close to \fp{} on CIFAR-10/Residual Network 18
(ResNet-18) and Stanford Sentiment Treebank 2 (SST-2)/DistilBERT, while
full-path low-bit aggregation fails on CIFAR-100/ResNet-18.  Layer-wise
diagnostics identify the classifier head as the sensitive component: keeping
the backbone low-bit while leaving the head on \fp{} recovers 73.0--73.1\%
accuracy versus 74.45\% for same-runner \fp{} and reduces gradient traffic to
3.6--5.4\% of the \fp{} baseline.  Hardware
synthesis and field-programmable gate array (FPGA) place-and-route estimates
suggest that the 512-bit aggregation datapath is small enough to be treated as a
near-memory datapath extension, not a separate accelerator-scale block.
\end{abstract}


\maketitle

\section{Introduction}
\label{sec:intro}

Large model training repeatedly moves gradient tensors through the memory and
network hierarchy.  For Generative Pre-trained Transformer 3
(GPT-3)-scale models~\cite{brown2020gpt3}, a full 32-bit gradient tensor is
hundreds of gigabytes; even smaller models can exceed graphics processing unit
(GPU) last-level cache (LLC) capacity during collective communication.  Once active
gradient footprints miss in cache, the memory-controller path becomes part of the
All-Reduce cost.

This regime is the reason to look below the usual software collective layer.
For small models, shaving gradient payloads at a Compute Express Link (CXL) controller is unlikely to
matter unless the experiment deliberately exposes cache misses and controller
service time.  For large training jobs, however, the same gradient movement is
repeated over many layers, buckets, and steps, so controller-side gradient
aggregation can be valuable even if the primitive is narrow.  In this paper, aggregation means combining gradient contributions
from multiple workers for the same parameter bucket. We
therefore use controlled vision and language fine-tuning runs to test
correctness and convergence boundaries, while using transformer-scale trace
replay to ask whether the datapath would be exposed in the communication path.

At the same time, sign-based gradient methods and low-bit learning systems such as BNN and BitNet show that many workloads can tolerate sign or ternary update
information~\cite{bernstein2018signsgd,hubara2016bnn,ma2024bitnet}.  This
motivates \sys{}, a CXL-side controller architecture for such low-bit gradient
aggregation. \sys{} does not change
model computation, model weights, or
backpropagation; it changes how gradient communication payloads are represented
and combined at the controller response path.

We call the two evaluated \sys{} controller primitives gradient-binary
(\gbin{}) and gradient-ternary (\gter{}).  The
arithmetic is borrowed from BNN and BitNet-style low-bit learning, but the controller
applies it only to gradient communication payloads.  \gbin{} reduces
each worker contribution to a sign-coded payload that maps naturally to
PopCount and narrow add-tree logic.  \gter{} keeps the same sign-count structure
but adds a zero gate that can suppress weak or uncertain entries.  These are
intentionally aggressive low-bit endpoints in the gradient-aggregation design
space: simple enough to fit in a controller-resident datapath, but not assumed
to be universally safe.  INT8, FP8, block-floating-point, or mixed-precision
formats could provide more conservative accuracy/traffic tradeoffs, at the cost
of scale metadata, wider arithmetic, and rounding or saturation policy.
\sys{} therefore treats precision as a policy-selectable dimension.

This naming is intentional.  \gbin{} and \gter{} are not claims that the model is
trained as a BNN or BitNet, nor that one-bit or ternary communication is always
the right accuracy point.  They name two gradient-payload formats that are
aggressive enough to stress the accuracy boundary and simple enough to evaluate
as controller-resident logic.  More conservative formats are compatible with the
same admission framework, but would answer a different cost/accuracy question.

Existing software compression and in-network aggregation systems cut
network/software traffic in the
stack~\cite{seide2014,lin2018dgc,sharp2016,switchml2021}; however, they do not answer a
different architectural question: can the Compute Express Link (CXL)-side memory controller aggregate
packed gradient signs as gradient cache lines pass through the controller?

\sys{} explores this design point.  The architecture places a
short \gbin{}/\gter{} aggregation datapath in the CXL controller and exposes a
control interface
that can keep \fp{} as the calibration/recovery path while admitting low-bit
aggregation only when the workload, layer, or phase can tolerate it.  The
central result is not that low-bit aggregation should always replace \fp{}, but
that a CXL controller can expose a hardware-light, traffic-saving aggregation
mode whose safe operating region can be identified by workload- and layer-aware
admission.
\sys{} is therefore not a universal replacement for
\fp{} All-Reduce, NVIDIA Collective Communications Library (NCCL)~\cite{nvidia_nccl}
/ Remote Direct Memory Access (RDMA), or switch aggregation.  Its value is to cut
admitted gradient traffic at the memory-controller boundary down to
3.6--5.4\% of the \fp{} baseline in the layer-aware
CIFAR-100~\cite{krizhevsky2009cifar} setting, while
retaining \fp{} as the calibration and recovery path when low-bit aggregation
would harm training.

\paragraph{Contributions.}
\begin{itemize}
  \item A CXL-side low-bit aggregation architecture that combines identity
        read-back, \gbin{} sign-count aggregation, and \gter{} ternary gating in the
        controller datapath.
  \item A timing and operating-envelope evaluation showing that the
        five-cycle low-bit aggregation datapath is not the bottleneck under the configured CXL
        service model and adds at most 1.67\% exposed overhead in the measured
        full LLC-miss regime.
  \item Evidence for accurate low-bit regimes: CIFAR-10 with
        ResNet-18~\cite{he2016resnet} and
        SST-2~\cite{socher2013sst,wang2018glue} with
        DistilBERT~\cite{sanh2019distilbert} retain near-\fp{} accuracy under
        selected low-bit aggregation settings.
  \item A boundary result and recovery mechanism: CIFAR-100 rejects full-path
        low-bit aggregation, but layer-aware admission recovers most accuracy
        by keeping the classifier head on \fp{} while reducing traffic by
        94.6--96.4\%.
  \item Hardware plausibility evidence from gate-equivalent modeling,
        Yosys~\cite{wolf2013yosys} synthesis, and nextpnr ECP5 routing
        checks~\cite{nextpnr,projecttrellis}.
\end{itemize}

\paragraph{Evidence scope.}
Our evaluation separates controller mechanism, training quality, scale-proxy,
and hardware-plausibility evidence. gem5 timing and the functional tester
validate the controller path and byte-level \gbin{}/\gter{} semantics; CIFAR-10,
SST-2, and CIFAR-100 test convergence regimes and boundary behavior. GPT-scale
trace replay models only the gradient-communication component, while synthesis
and FPGA proxies bound datapath plausibility rather than replacing foundry
implementation.

The rest of the paper is organized as follows. Section~\ref{sec:background}
defines the CXL near-memory opportunity, the low-bit aggregation semantics, and
the scope of the sign-gradient baselines. Section~\ref{sec:arch} presents the
\sys{} architecture, including the controller datapath, the \gbin{}/\gter{}
aggregation datapath, and the \fp{} calibration and recovery path.
Section~\ref{sec:methodology} describes the simulation, functional, training,
and hardware-evaluation methodology. Section~\ref{sec:timing} quantifies the
timing overhead and operating envelope of the controller datapath, and
Section~\ref{sec:functional} validates the packed-sign aggregation semantics.
Section~\ref{sec:training} evaluates training accuracy, traffic reduction, and
the workload boundary exposed by CIFAR-100. Section~\ref{sec:control} reports
control-plane pilot results, Section~\ref{sec:positioning} positions \sys{}
against software and optimizer baselines, and Section~\ref{sec:hardware}
summarizes synthesis and place-and-route evidence. Finally,
Section~\ref{sec:limitations} discusses limitations and future work, and
Section~\ref{sec:conclusion} concludes.

\section{Background}
\label{sec:background}
This section defines the CXL placement, low-bit aggregation semantics, and
baseline taxonomy used by the rest of the paper.
\paragraph{CXL near-memory opportunity.}
CXL 3.0 provides coherent host/accelerator access to
memory pools and fabric devices~\cite{cxl30spec}.  Prior CXL memory work
focuses on pooling, disaggregation, or latency/bandwidth
management~\cite{pond2023cxl,gouk2022cxl,li2023cxlmem}.  \sys{} instead asks
whether a CXL controller can perform useful gradient aggregation while serving
CXL-targeted cache-line traffic.

This is narrower than general near-memory computation.  The operation is not
arbitrary tensor execution near memory; it is a fixed cache-line reduction over
a small packed representation.  That restriction is important for hardware
plausibility.  A CXL memory controller is not a GPU, and the design should not
depend on large static random-access memory (SRAM) buffers, floating-point
datapaths, or software-managed compute kernels inside the controller.  The opportunity
is that gradient communication already moves similarly structured cache lines
through the controller, and sign or ternary aggregation can be implemented with
regular bitwise logic.

\paragraph{Low-bit aggregation semantics.}
The low-bit signal helps because gradient communication is often bandwidth-bound.
Replacing \fp{} gradient values with packed signs or ternary symbols reduces the
gradient payload, and the corresponding aggregation can be implemented with
regular bitwise logic rather than floating-point addition.

Let $W$ be the number of workers, $k$ index a worker, and $i$ index a gradient
element.  Worker $k$ still produces an ordinary \fp{} gradient value
$g_{k,i}$; for an admitted low-bit bucket, the communication runtime
encodes its sign as $b_{k,i}$ before writing the CXL-resident payload.
The controller consumes these packed bits, where $b_{k,i}=1$ denotes a positive
sign and $b_{k,i}=0$ denotes a non-positive sign, and forms:
\[
  \begin{aligned}
  b_{k,i} &= \mathbf{1}\{\operatorname{sgn}(g_{k,i}) > 0\},\\
  c_i &= \operatorname{PopCount}(b_{0,i},\ldots,b_{W-1,i}),\\
  a_i &= 2c_i-W,\\
  u_i^{\mathrm{bin}} &= \operatorname{sgn}(a_i).
  \end{aligned}
\]
Here $c_i\in[0,W]$ is the sign count and $a_i\in[-W,W]$ is the signed vote
margin.  The returned \gbin{} update is $u_i^{\mathrm{bin}}$.  The packed functional
test checks the byte-level count $c_i$, while the convergence tests interpret
the same count through $u_i^{\mathrm{bin}}$.

\gter{} adds a zero gate $m_i\in\{0,1\}$ and returns
\[
  u_i^{\mathrm{ter}}=m_i u_i^{\mathrm{bin}} .
\]
Unless otherwise stated, the evaluated controller gate is a fixed 2-of-3 pattern
over flattened gradient elements: two consecutive elements keep the \gbin{} update,
and the third returns zero.  For example, with eight workers, a six-positive,
two-negative vote returns a positive \gbin{} direction; if the \gter{} gate is zero
for that element, the \gter{} datapath returns zero instead.  

In this paper,
\gbin{} and \gter{} denote controller-side communication primitives inspired by
BNN/BitNet arithmetic; they do not quantize the model weights or claim the full
co-designed training recipe of prior low-bit model systems~\cite{ma2024bitnet}.

This point matters for evaluation. Prior low-bit model systems can use
model-aware recipes such as scaling, clipping, optimizer changes, and
layer-specific training schedules. \sys{} starts from a more constrained
substrate: the controller observes the gradient communication payload and can
replace the returned aggregate with a low-bit signal. This constrained placement
also lowers the integration burden. Unlike full low-bit model recipes, \sys{}
does not require rewriting the model architecture, changing layer definitions,
or retraining with a model-specific quantization schedule. The runtime
encodes compact gradient payloads, and the controller aggregates them with
regular bitwise/counting logic. The tradeoff is that this hardware-light path
cannot rely on model-aware compensation; it removes magnitude information and
therefore places more burden on admission control.

Admission control is the policy that decides whether a
workload, layer group, or training phase is allowed to use the low-bit
aggregation path.  If diagnostics indicate that the low-bit signal is reliable,
the controller admits \gbin{} or \gter{} aggregation.  If not, the controller keeps
that group on the \fp{} path for calibration or recovery. Later experiments evaluate when each workload and layer group can safely use
low-bit aggregation and when the controller should remain on the \fp{} path.
\paragraph{Sign-gradient baselines.}
We include two sign-gradient baselines because a ``sign'' update can be formed
at different points in the reduction pipeline. MajoritySignSGD is
communication-comparable to \gbin{}: each worker contributes only a sign for
each gradient element, and the update direction is determined by the majority
sign across workers. It therefore tests whether sign-only distributed reduction
can preserve accuracy, although it is still implemented as a software baseline
rather than a CXL datapath.

SignOfMean is a stronger optimizer reference but is not communication-comparable.
It first computes the \fp{} mean gradient and only then takes the sign, so the
full-precision reduction has already been performed. We include it to separate
the convergence strength of sign updates from the hardware question of whether
a controller-resident low-bit aggregation primitive can provide the same
communication savings.

\paragraph{Evaluation assumptions.}
Four modeling choices bound what the evaluation claims. First, gem5
\cite{binkert2011gem5,lowepower2020gem5} timing uses configured DDR5-like DRAM
and CXL service points plus a five-cycle aggregation datapath, meaning a
five-controller-cycle pipeline for the low-bit block; these are modeling points,
not measured commercial-device numbers. Second, gem5 timing experiments retain
ordinary identity read-back, while a separate functional test validates the
\gbin{} and \gter{} transformed-payload semantics. Third, because gem5,
Astra-Sim~\cite{astrasim2020}, and NS-3~\cite{ns3} model different clocks and
levels of detail, collective timing and network telemetry are passed into gem5
as traces rather than through synchronized co-simulation. Fourth, convergence
experiments split minibatches into virtual workers to test aggregation semantics,
not eight-physical-GPU NCCL scaling.

Table~\ref{tab:requirements} summarizes the resulting design requirements and points to
the evidence used to validate each one.

\begin{table}[!t]
  \centering
  \caption{Design requirements and validation evidence.}
  \label{tab:requirements}
  \tablebodysetup
  \begin{tabularx}{\columnwidth}{@{}YYY@{}}
    \toprule
    Requirement & Reason & Evidence \\
    \midrule
    Preserve \fp{} path & needed for calibration and recovery & control pilots \\
    Keep datapath short & must fit CXL service interval & timing envelope \\
    Validate byte semantics & low-bit path must not corrupt read-back & functional test \\
    Admit selectively & low-bit is not safe for all workloads/layers & CIFAR-100 boundary \\
    Keep hardware regular & controller logic must scale with line width & synthesis/routing \\
    \bottomrule
  \end{tabularx}
\end{table}

\section{Architecture}
\label{sec:arch}

The previous section defined the low-bit payload semantics and the evaluation
assumptions. This section turns those semantics into a controller interface:
a read-response datapath for admitted CXL-resident gradient payloads, plus a
control path that selects whether the response is transformed by the low-bit
datapath or left on the normal full-precision route.

\begin{figure}[!t]
  \centering
  \begin{tikzpicture}[
    font=\scriptsize,
    block/.style={
      draw, rounded corners=1.5pt, align=center, fill=white,
      minimum height=7.5mm, inner sep=2.5pt, line width=0.45pt
    },
    datapath/.style={-{Stealth[length=1.9mm]}, line width=0.52pt},
    ctrlpath/.style={-{Stealth[length=1.9mm]}, dashed, line width=0.5pt},
    lowbit/.style={block, fill=blue!6, text width=1.9cm},
    bypass/.style={block, fill=orange!10, text width=2.25cm},
    meta/.style={block, fill=gray!10, text width=3.1cm}
  ]
    \node[block, text width=1.0cm] (workers) at (-0.1,0)
      {GPU\\workers};
    \node[block, text width=1.05cm] (llc) at (1.25,0)
      {shared LLC\\cache lines};
    \node[block, text width=1.05cm] (switch) at (2.6,0)
      {CXL\\fabric/link};

    \node[draw, rounded corners=2pt, minimum width=2.85cm, minimum height=2.15cm,
      inner sep=0pt, line width=0.45pt] (controller) at (4.75,0) {};
    \node[anchor=south, font=\scriptsize\bfseries] at
      ([yshift=1pt]controller.north) {CXL controller};

    \node[block, fill=gray!10, text width=1.45cm, minimum height=5.5mm]
      (latch) at (4.75,0.5) {mode latch};
    \node[block, fill=blue!6, text width=1.1cm, minimum height=8.5mm]
      (lane) at (4.08,-0.42) {\gbin{}/\\\gter{}\\datapath};
    \node[block, fill=orange!8, text width=0.95cm, minimum height=7.5mm]
      (bypass) at (5.38,-0.42) {bypass\\return};

    \node[block, text width=1.05cm] (memory) at (7.05,0)
      {pooled\\CXL memory};
    \node[bypass] (fp32) at (1.95,-1.75)
      {\fp{} runtime\\aggregation\\software fallback};

    \node[meta] (control) at (4.75,2.05)
      {Control path:\\trace forecast + fabric telemetry\\loss feedback};

    \draw[datapath] (workers) -- (llc);
    \draw[datapath] (llc) -- (switch);
    \draw[datapath] (switch) -- (controller.west);
    \draw[datapath] (controller.east) -- (memory);
    \node[font=\tiny, align=center] at (2.25,0.58)
      {write-side\\payload};
    \draw[datapath] (memory.south west) to[out=-160,in=-20]
      node[below, yshift=-14pt, font=\tiny, align=center, fill=white, inner sep=0.8pt]
      {read-response\\transform}
      (controller.south east);
    \draw[datapath] (workers.south) to[out=-90,in=180] (fp32.west);
    \draw[datapath] (fp32.north) to[out=90,in=-90] (llc.south);

    \draw[ctrlpath] (control.south) -- (latch.north);
    \draw[ctrlpath] (latch.south) -- (lane.north);
    \draw[ctrlpath] (latch.south) -- (bypass.north);

    \draw[datapath] (-0.55,1.25) -- ++(0.45,0)
      node[right, font=\tiny, inner sep=1pt] {payload};
    \draw[ctrlpath] (-0.55,1.03) -- ++(0.45,0)
      node[right, font=\tiny, inner sep=1pt] {mode metadata};
  \end{tikzpicture}
  \caption{\sys{} node architecture.  The runtime writes ordinary \fp{} or
  admitted packed low-bit payloads into the CXL-resident gradient buffer; on a
  later read response, the controller selects identity/bypass or the
  \gbin{}/\gter{} aggregate transform without overwriting stored bytes.  A
  discrete CXL switch is only required in switched pooled-memory deployments.}
    \Description{Architecture diagram showing GPU workers, shared cache, CXL fabric
    link, CXL memory controller, mode latch, \gbin{} and \gter{} low-bit aggregation datapath, pooled CXL
    memory, control modules, and a separate full-precision runtime fallback path.}
\label{fig:arch}
\end{figure}
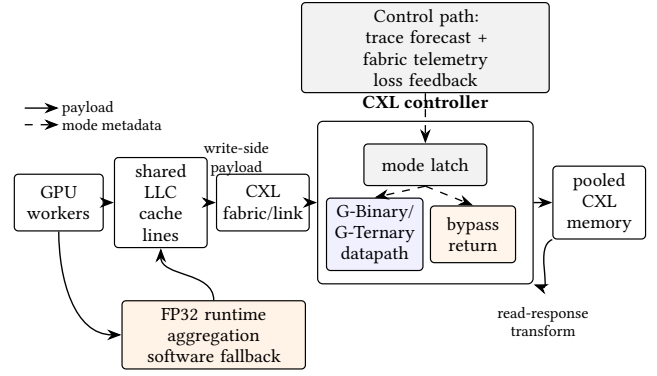

Fig.~\ref{fig:arch} shows the node-level organization.  GPU gradient traffic
targets a CXL-attached memory region.  The datapath carries cache-line payloads
and returns either identity bytes, an \fp{} bypass response, or a
\gbin{}/\gter{} aggregate; the control path only writes mode metadata. For an admitted low-bit bucket, the runtime derives a communication payload from
the ordinary \fp{} gradients before the write: \gbin{} stores one sign bit per
gradient element, while \gter{} stores sign bits plus zero-gate bits. The model
and backpropagation still produce \fp{} gradients; only the communication payload
is encoded into a packed low-bit format. Runtime then writes that
packed payload into the CXL-resident gradient buffer. On the later read response, the
controller applies the selected identity, \gbin{}, \gter{}, or \fp{} path.

\paragraph{CXL protocol scope.}
\sys{} places aggregation on the CXL.mem-side response path for gradient
buffers.  CXL.io handles ordinary enumeration/configuration, and CXL.cache is
unchanged in our model.  Thus \sys{} changes the returned aggregation payload,
not the cache-line or coherence protocol.  The controller does not run
backpropagation, optimizer logic, or model-level quantization; those remain in
the training stack.  It only transforms admitted gradient communication
payloads on the response path.

Table~\ref{tab:modes} defines the four payload modes selected by the control
path.

\begin{table}[!t]
  \centering
  \caption{Controller-visible aggregation modes selected by the control path.}
  \label{tab:modes}
  \tablebodysetup
  \begin{tabularx}{\columnwidth}{@{}lYY@{}}
    \toprule
    Mode & Returned payload & Role in the policy \\
    \midrule
    identity & original cache-line bytes & functional read-back check \\
    \fp{} bypass & normal full-precision path & warm-up, calibration, recovery \\
    \gbin{} & majority sign aggregate & 1-bit low-traffic relief \\
    \gter{} & ternary sign/zero aggregate & sparse low-bit relief \\
    \bottomrule
  \end{tabularx}
\end{table}

\paragraph{Datapath.}
The low-bit datapath is 512 bits wide, matching the 64-byte CXL response
granularity.  In \gbin{} mode, each byte encodes packed signs from eight virtual
workers and XNOR/PopCount logic computes the returned sign count.  \gter{}
reuses the count datapath and applies the recurring ternary zero gate.  The
modeled five-cycle delay covers request decode, sign unpacking/alignment,
per-element counts, majority or ternary gating, mode selection, and response
registration; Section~\ref{sec:hardware} checks that this depth is plausible
for a 2\,GHz target rather than assuming a single-cycle 512-bit block.

\paragraph{Control interface.}
The controller exposes a mode latch rather than embedding training policy in
the datapath. The control plane is organized into three roles: a Predictor that
estimates collective pressure from trace-derived timing and bandwidth forecasts,
a Commander that proposes a mode from network telemetry, and a Supervisor that
keeps or recovers to \fp{} when training-health telemetry is unsafe. In this paper, network telemetry is replayed communication-state information
such as bandwidth pressure and cache-line demand, while training-health
telemetry is runtime feedback such as loss trend, warm-up diagnostics, and
recovery triggers. These signals drive mode selection; the controller itself
receives only mode metadata. The
Predictor does not observe gradients, weights, or loss; its stored forecasts
include forward/backward duration, All-Reduce timing, gradient volume, shard
bytes per GPU, peak CXL bandwidth demand, and cache-line read count. The control
plane writes only mode metadata; it does not inspect or rewrite gradient
payloads.

\paragraph{Admission and recovery.}
Training begins on the \fp{} bypass path while the control plane records loss,
gradient, and communication telemetry.  If the Commander admits a low-bit mode for a phase or parameter group, it writes
the corresponding \gbin{} or \gter{} mode from Table~\ref{tab:modes} into the
controller mode latch.  Later reads still follow the normal CXL
read/response flow, but the controller returns the sign-count aggregate instead
of the bypass payload.  If the Supervisor observes sustained loss growth or
another recovery trigger, it clears the latch back to \fp{} for cooldown and
calibration; after recovery, the same admission rule may re-enable low-bit
aggregation.

\paragraph{Write-side payload materialization.}
\sys{} keeps the training interface and the memory-controller interface
separate. Model forward/backward computation still produces ordinary \fp{}
gradients, and the training stack still decides which gradient buckets are being
communicated. For an admitted low-bit bucket, the communication runtime derives
a compact communication representation from those \fp{} gradients before the
write: \gbin{} uses packed sign bits, while \gter{} uses packed sign bits plus
zero-gate bits. The runtime then writes that representation into the
CXL-resident gradient buffer as ordinary cache-line payloads. In other words,
the CXL-resident region remains a byte-addressed gradient communication buffer.

\paragraph{Read-side aggregation.}
The controller-side operation happens later, when an aggregate is read. The
memory controller does not interpret model structure, tensor semantics, layer
roles, or training state, nor does it execute backpropagation or optimizer
logic. The controller operates at CXL cache-line granularity. For an admitted
aggregate read, it interprets the returned cache-line payload under the selected
mode. The read-side transform is non-destructive: it does not overwrite the
bytes at the target addresses or mutate model/optimizer state. It only changes
what the requester receives on that aggregate read response. Warm-up, rejected
admission, calibration, and recovery use the ordinary \fp{} payload path, so the
same buffer interface can fall back to full-precision communication when low-bit
aggregation is unsafe.

\paragraph{Latency decomposition.}
Let $T_\text{agg}$ denote the low-bit aggregation delay.  The timing model
includes a per-line CXL bandwidth gate: a configurable limiter that spaces
cache-line service events according to the available link bandwidth.  We call
this spacing the bandwidth-gate interval.  Because aggregation is registered
behind the same controller service, that interval can hide part or all of the
low-bit aggregation datapath.  The exposed low-bit component is approximately
\[
  T_\text{exposed} = \max(0, T_\text{agg} - T_\text{overlap}),
\]
where $T_\text{overlap}$ is the part of the bandwidth-gate interval and
adjacent queued memory service that overlaps the aggregation datapath.  This is why
\sys{} is useful only in some regimes: the same five-cycle datapath can be hidden
under bandwidth pressure or exposed when controller queues are sparse.

\section{Methodology}
\label{sec:methodology}

The methodology follows the architectural decomposition above. It proceeds from
controller timing and byte semantics to training quality, admission, control,
and hardware plausibility: first isolating when the read-side low-bit transform
is exposed by the CXL cache-line service path, then separating that timing
question from correctness, convergence, control behavior, and implementation
plausibility. Table~\ref{tab:evidence-map} summarizes these evidence blocks.

The timing experiments use gem5 with controlled DDR5-like and CXL-like service
parameters. Astra-Sim supplies collective timing for the large language model
(LLM)-scale projection, reported using BERT-large, GPT-2 XL, and GPT-3 model
profiles. NS-3 supplies telemetry replay for the control-path check. The
training experiments run model compute on Compute Unified Device Architecture
(CUDA) where applicable and apply the selected aggregation rule to
virtual-worker gradients, following the evaluation assumptions in
Section~\ref{sec:background}.

All accuracy numbers are reported as top-1 validation accuracy for vision or
classification accuracy for SST-2.  Communication ratios count the gradient
payload transferred under the selected aggregation representation and normalize
it to the same-runner \fp{} gradient payload.  The ratios are therefore not
wall-clock training speedups; they are the communication pressure that the
controller-side mechanism would remove if the selected low-bit mode is
admitted.
Unless otherwise stated, low-bit traffic results assume that admitted buckets
have already been encoded by the communication runtime as the selected packed sign
or ternary payloads before entering the CXL-resident buffer.

\begin{table*}[!t]
  \centering
  \caption{Evaluation evidence map and workload scope.}
  \label{tab:evidence-map}
  \tablebodysetup
  \begin{tabularx}{\textwidth}{@{}>{\raggedright\arraybackslash}p{0.15\textwidth}
                  >{\raggedright\arraybackslash}p{0.24\textwidth}
                  >{\raggedright\arraybackslash}p{0.22\textwidth}
                  Y@{}}
    \toprule
    Claim & Evidence block & Workload/protocol & What it answers \\
    \midrule
    Timing exposure & gem5 timing, bucket replay, operating-envelope sweep &
    synthetic cache-line streams; BERT/GPT profiles &
    whether the datapath is exposed in the controller path and whether bursty
    gradient buckets change that conclusion \\
    Functional correctness & packed-sign validation &
    packed eight-worker sign packets; deterministic controller test &
    whether identity, \gbin{}, and \gter{} read-back semantics are exact \\
    Validated convergence & CIFAR-10 and SST-2 runs &
    ResNet-18 and DistilBERT; CUDA training/fine-tuning with 8 virtual workers &
    whether low-bit aggregation works in validated vision and Transformer
    regimes \\
    Boundary/admission & CIFAR-100 and layer-aware runs &
    ResNet-18; CUDA training with 8 virtual workers &
    where full-path low-bit aggregation fails and how the failure localizes into
    an admission rule \\
    Control safety & CUSUM and guarded recovery pilots &
    CIFAR-10/ResNet-18 traces with live mode selection and controlled
    degradation windows &
    whether telemetry and training feedback can keep \fp{} available for
    warm-up, rejection, and recovery \\
    Hardware plausibility & datapath synthesis and place-and-route summary &
    512-bit \gbin{}/\gter{} datapath cost model &
    whether the datapath is small and regular enough for a controller-resident
    implementation \\
    \bottomrule
  \end{tabularx}
\end{table*}

\paragraph{Default configuration.}
Unless otherwise stated, gem5 timing runs use eight virtual workers issuing
64-byte cache-line accesses through a configured CXL service model with
128\,GiB/s bandwidth, 200\,ns fixed memory-access latency, and an 8\,MiB LLC.
The controller datapath uses the 512-bit, five-controller-cycle \gbin{}/\gter{} datapath
from Section~\ref{sec:arch}.  The measured exposed-cost anchor uses a
2\,MiB/GPU active gradient-buffer footprint across the eight workers, forcing
the 16\,MiB active footprint beyond the LLC; trace replay and the envelope sweep
inherit this full-LLC-miss anchor.  CUDA convergence experiments run on one CUDA
device and simulate eight workers by splitting each batch before software
aggregation; they check convergence and communication representation, not
physical multi-GPU throughput.

\section{Low-Bit Datapath Timing Envelope}
\label{sec:timing}

This section separates controller-datapath cost from memory service, cache
filtering, bucket scheduling, and controller scale-out.  The goal is to bound
when the five-cycle low-bit datapath becomes visible, not to claim that low-bit
aggregation always accelerates a step.

\subsection{Measured Datapath Exposure}

Table~\ref{tab:timing-summary} reports the exposed cost of adding the five-cycle
\gbin{} datapath, i.e., the portion not hidden by memory-service and
bandwidth-gate delay. We refer to this non-congested service setting as
healthy-CXL; in congested-CXL settings, bandwidth-gate and queueing delay can
hide the low-bit work. Under the default
modeling anchor in Section~\ref{sec:methodology}, \gbin{} adds 1.11\% overhead with 
one synthetic cache-line request stream and remains below 2\% across measured scaling and
full-LLC-miss runs. These numbers isolate the incremental cost of the low-bit datapath under the
configured service model; they do not compare commercial CXL and DDR5 DRAM memory
systems.

\begingroup
\setlength{\intextsep}{0.25\baselineskip}
\begin{table}[!t]
  \centering
  \caption{Measured \gbin{} datapath timing exposure.}
  \label{tab:timing-summary}
  \tablebodysetup
  \begin{tabularx}{\columnwidth}{@{}YYY@{}}
    \toprule
    Result & Observation & Takeaway \\
    \midrule
    1-request stream \gbin{} overhead & 1.11\% & low exposed cost \\
    1--8 request stream overhead & 0.55--1.88\% & bounded scaling \\
    full LLC-miss overhead & 1.67\% & worst measured exposure \\
    scale-out, healthy & 1.07--1.67\% & bounded across 1--8 controller stacks \\
    scale-out, congested & hidden--1.11\% & overlap can disappear as stacks split \\
    \bottomrule
  \end{tabularx}
\end{table}
\endgroup

A GPT-scale sanity check multiplies Astra-Sim All-Reduce windows for
BERT-large~\cite{devlin2019bert}, GPT-2 XL~\cite{radford2019gpt2}, and
GPT-3~\cite{brown2020gpt3} by the 1.67\% full-LLC-miss component, giving 0.78M,
3.4M, and 399M exposed cycles.  The workload-shaped replay and sensitivity
sweep provide the main timing evidence.

\subsection{Bucketized Trace Replay}

\begin{figure}[!t]
  \centering
  \includegraphics[width=\linewidth]{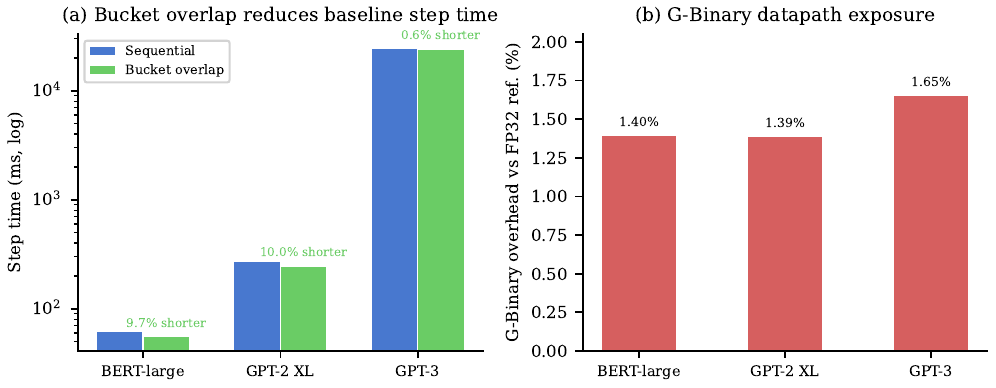}
  \caption{Gradient-bucket trace replay.  Panel (a) shows baseline bucket
  overlap with the backward tail.  Panel (b) isolates healthy-CXL exposed cost
  after adding the \gbin{} datapath; congested-CXL cases are hidden by the
  bandwidth gate.}
    \Description{Trace replay plot showing bucket-overlap scheduling effects and \gbin{} low-bit path exposure under healthy CXL.}
\label{fig:trace-replay}
\end{figure}

The trace replay replaces the uniform stream with 32\,MiB gradient buckets from
BERT-large, GPT-2 XL, and GPT-3 profiles.  Bucket overlap is a baseline
framework scheduling effect: early buckets communicate while later backward
operators still run.  Fig.~\ref{fig:trace-replay} asks whether that burstiness
exposes more controller cost.  In the healthy-CXL case, adding the \gbin{}
datapath leaves only a 1.25--1.65\% increment over the \fp{} no-aggregation
reference.

\subsection{Envelope and Scale-Out}

\begin{figure*}[!t]
  \centering
  \includegraphics[width=0.95\textwidth]{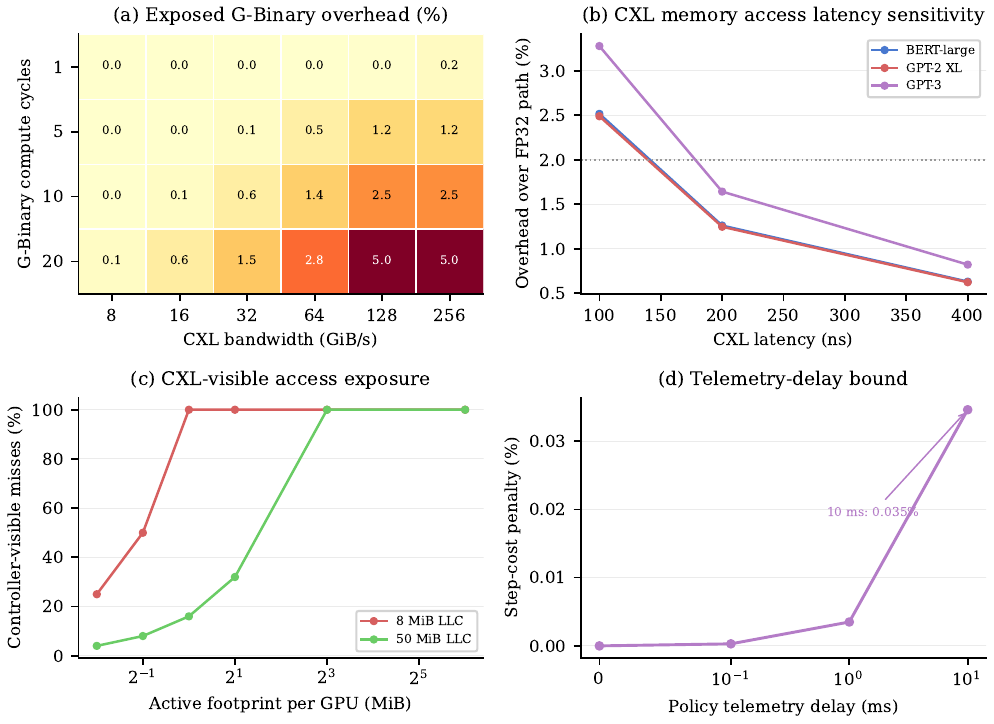}
  \caption{Operating-envelope sensitivity sweep.  Panel (a) maps exposed
  overhead across CXL bandwidth and G-Binary compute cycles; panel (b) varies the
  fixed CXL memory-access latency; panel (c) shows how LLC capacity controls
  controller-visible accesses; panel (d) bounds policy telemetry delay as a
  late mode update.}
    \Description{Multi-panel sensitivity sweep showing when bandwidth, delay, latency, CXL-visible access exposure, and telemetry delay make the low-bit stage visible.}
\label{fig:sensitivity-envelope}
\end{figure*}

Fig.~\ref{fig:sensitivity-envelope} maps when the timing conclusion stops
holding. Panel~(a) varies CXL bandwidth and low-bit datapath delay. The measured
5-cycle datapath stays below a 2\% exposed-cost gate across the bandwidth sweep,
while deeper 10-cycle and 20-cycle datapaths exceed the gate at high CXL
bandwidth because less bandwidth-gate delay is available to hide computation.
Panel~(b) varies the fixed CXL memory-access latency; changing this latency
shifts the base memory-service path but does not make the measured 5-cycle
datapath the dominant cost. Panel~(c) varies the active gradient footprint
relative to LLC capacity and shows why the full-LLC-miss point is the
conservative exposed case: once the active footprint exceeds LLC capacity, more
accesses reach the CXL controller. Panel~(d) delays mode telemetry updates and
shows that even a 10\,ms late update adds less than 0.04\% step cost in this
model.

The scale-out rows in Table~\ref{tab:timing-summary} add one more stress case. Splitting one shared CXL controller
into $N\in\{1,2,4,8\}$ stacks reduces queueing but can also remove overlap that
previously hid the low-bit datapath. Across the 8-GPU LLC-miss sweep, the
exposed overhead remains bounded rather than growing with stack count.

\section{Functional Correctness}
\label{sec:functional}

The timing model is meaningful only if the controller implements the intended
aggregation semantics.  The subtlety is the validation oracle.  A conventional
memory regression assumes identity memory and would flag any read value that
differs from the last write to the same address as a failure.  That check
remains correct for identity mode, but it is the wrong oracle for \gbin{} and \gter{}
because their read response is intentionally a transformed aggregate over the
worker packets.  We therefore validate the controller with mode-specific
expected values: identity mode uses byte-for-byte read-back, while \gbin{} and
\gter{} use a transformation-aware software oracle that computes the
Section~\ref{sec:background} reduction before comparing against the controller
response.

The packed-sign validation writes sign packets for eight virtual workers and
reads them back under three modes: identity, \gbin{}, and \gter{}.  This is a fixed
64-byte-packet read-back test over the same address range, not a traffic-volume
measurement.  The harness selects the expected payload from the active
controller mode rather than forcing every mode to satisfy read-equals-write.
The payloads returned by all three modes match their corresponding oracle
exactly, closing the byte-semantics check before the paper turns to training
quality.

\section{Training Evidence and Boundary}
\label{sec:training}

Functional correctness does not imply training quality.  This section first
checks where \gbin{}/\gter{} preserve convergence, then uses a harder workload
to expose the admission boundary.

\subsection{Validated Low-Bit Regimes}

\begin{figure*}[!t]
  \centering
  \includegraphics[width=0.47\textwidth]{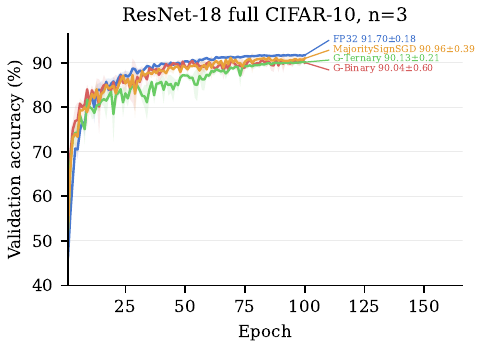}\hfill
  \includegraphics[width=0.47\textwidth]{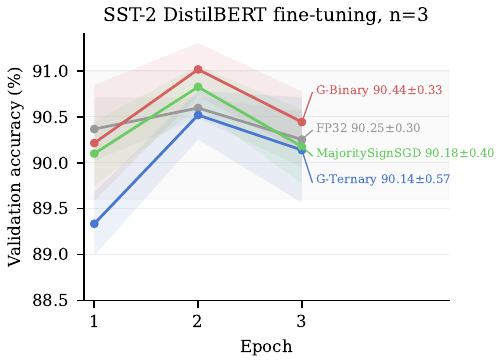}
  \caption{Validated low-bit convergence regimes.  Left: full 100-epoch
  CIFAR-10/ResNet-18 training.  Right: complete three-epoch SST-2/DistilBERT
  fine-tuning.  Endpoint labels report final mean $\pm$ standard deviation
  across three seeds.}
    \Description{Two convergence plots comparing FP32, \gbin{}, \gter{}, and MajoritySignSGD on CIFAR-10 and SST-2.}
\label{fig:validated-convergence}
\end{figure*}

Fig.~\ref{fig:validated-convergence} combines convergence curves with final
mean and standard deviation at each endpoint.  On CIFAR-10/ResNet-18,
\gbin{} and \gter{} remain within 2 points of \fp{} and within 1 point of
MajoritySignSGD.  On full-data SST-2/DistilBERT, all three low-bit/sign
methods remain in the same final-accuracy band as \fp{} across three seeds.  The three SST-2 epochs are the full supervised fine-tuning schedule from a
pretrained DistilBERT checkpoint, not a truncated pretraining run.

\subsection{Harder-Workload Boundary}

\begin{figure}[!t]
  \centering
  \includegraphics[width=\linewidth]{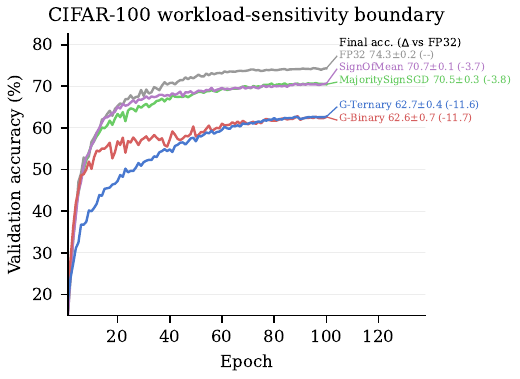}
  \caption{CIFAR-100 boundary.  Full-path \gbin{} and \gter{} remain far below
  \fp{} and the sign-gradient references.  Endpoint labels report final mean
  $\pm$ standard deviation and $\Delta$ versus \fp{} across three seeds.}
    \Description{CIFAR-100 accuracy plot showing full-path \gbin{} and \gter{} below FP32 and sign-gradient baselines.}
\label{fig:cifar100-boundary}
\end{figure}

We use boundary here in an empirical sense: it is the point where applying the
same low-bit aggregation policy to every layer no longer preserves training
quality. CIFAR-100/ResNet-18 exposes this case: Fig.~\ref{fig:cifar100-boundary}
shows that full-path \gbin{} and \gter{} lag \fp{} by about 11.6 points and trail
the sign-gradient references by about 8 points. This is a useful negative result rather than a failed run. It leads to the systems conclusion that admission must be workload- and
layer-aware.

CIFAR-100 keeps the
same architecture family as CIFAR-10 but makes the classification problem
harder, which separates ``the low-bit aggregation datapath can execute'' from ``the whole workload can tolerate removing magnitude information everywhere.''  A universal low-bit
path would be attractive for hardware simplicity, but this result shows that the
control interface must preserve a normal \fp{} route for sensitive workloads,
layers, or phases.

\subsection{Layer-Aware Admission}

Layer-wise diagnostics localize the failure.  We measure low-bit/\fp{} cosine
alignment between the low-bit and \fp{} aggregate vectors for the same layer
group; values near 1 indicate matching update directions, while values near 0
indicate a nearly orthogonal signal.  At epoch 20, Table~\ref{tab:layer-diagnostics}
shows weak CIFAR-100 classifier-head alignment but a well-aligned backbone,
suggesting a selective policy: low-bit backbone, \fp{} head.

\begin{table}[H]
  \centering
  \caption{Epoch-20 low-bit/\fp{} cosine diagnostics.}
  \label{tab:layer-diagnostics}
  \tablebodysetup
  \begin{tabular}{@{}lccc@{}}
    \toprule
    Dataset & Global & Backbone & Head \\
    \midrule
    CIFAR-10 \gbin{}/\gter{} & .505/.415 & .694/.567 & .531/.435 \\
    CIFAR-100 \gbin{}/\gter{} & .510/.419 & .723/.591 & .174/.142 \\
    \bottomrule
  \end{tabular}
\end{table}

Table~\ref{tab:layer-aware} validates the diagnostic.  Low-bit backbone with an
\fp{} head recovers most accuracy while preserving most traffic reduction:
\gbin{} reaches $73.11\pm0.11$\% at $0.0357\times$ \fp{} gradient traffic, and
\gter{} reaches $73.04\pm0.32$\% at $0.0537\times$.  The reverse split is weaker
and retains almost all \fp{} traffic, confirming that the classifier-head signal is the sensitive component.
Traffic ratios count only gradient payload representation, not unrelated
framework overheads or end-to-end training time.

\begin{table}[!t]
  \centering
  \caption{CIFAR-100 layer-aware mixed aggregation, three seeds.}
  \label{tab:layer-aware}
  \tablebodysetup
  \begin{tabularx}{\columnwidth}{@{}Yrr@{}}
    \toprule
    Policy & Final acc. (\%) & Traffic vs.\ \fp{} \\
    \midrule
    \fp{} all & $74.45\pm0.28$ & 1.0000 \\
    \gbin{} all & $62.33\pm0.31$ & 0.0313 \\
    \gter{} all & $62.22\pm0.64$ & 0.0494 \\
    \gbin{} backbone + \fp{} head & $73.11\pm0.11$ & 0.0357 \\
    \gter{} backbone + \fp{} head & $73.04\pm0.32$ & 0.0537 \\
    \fp{} backbone + \gbin{} head & $72.46\pm0.36$ & 0.9956 \\
    \fp{} backbone + \gter{} head & $67.26\pm0.30$ & 0.9957 \\
    \bottomrule
  \end{tabularx}
\end{table}

A final ablation checks whether the layer-aware recovery is merely a
learning-rate effect. Because the tuned low-bit and \fp{} runs use different
learning rates, we keep the classifier head on \fp{} aggregation but assign it
the low-bit learning rate. The model still reaches $73.24\pm0.56$\% with a
\gbin{} backbone and \fp{} head, and $73.22\pm0.31$\% with a \gter{} backbone and
\fp{} head. Recovery therefore comes primarily from preserving the head
aggregation signal rather than from the learning-rate choice.

The boundary lesson is not to disable low-bit aggregation entirely. Instead, the
controller should admit the large, well-aligned backbone while keeping the
sensitive head on \fp{}. The always-on low-bit execution lacks that calibration and
recovery point.

\section{Control-Plane Pilots}
\label{sec:control}

\sys{} exposes a control interface because training evidence rejects
unconditional full-path low-bit execution. The method therefore keeps \fp{} as a calibration and recovery path: training begins on \fp{}, low-bit
modes are admitted only after diagnostics pass, and unsafe buckets or layer groups can return to
\fp{} when telemetry or training-health guards fail. The pilot question is
whether telemetry and training-health feedback can drive these online mode
choices.

For this paper, the ladder is not a learned policy. It maps diagnostics to the lowest-traffic
mode that passes, keeps a sensitive head on \fp{} when only the backbone passes,
and recovers to \fp{} if training-health telemetry degrades. The
calibration remains workload-specific while adaptive Commander/Supervisor policies
are future work.

Fig.~\ref{fig:guarded-recovery} expands one CIFAR-10/ResNet-18 recovery pilot
into an epoch-level trace. It uses four traces to separate references from
control behavior: always-\fp{} and always-\gbin{} are fixed-mode baselines,
\fp{}-default tests admission into low-bit aggregation, and \gbin{}-default
tests \fp{} recovery from low-bit execution.  After admission, the Supervisor
can return to \fp{} during an injected degradation window and re-enable low-bit
aggregation after recovery.  The \gbin{}-default policy keeps 81.11\% of steps
low-bit, uses 0.214x \fp{} communication on average, and ends at
$90.87\pm1.04$\%. The endpoint stays within the run-to-run variation of the \fp{} and
always-\gbin{} references, so the pilot demonstrates guarded recovery behavior
rather than a fixed universal threshold.

\begingroup
\setlength{\intextsep}{0.45\baselineskip}
\begin{figure}[H]
  \centering
  \includegraphics[width=\linewidth]{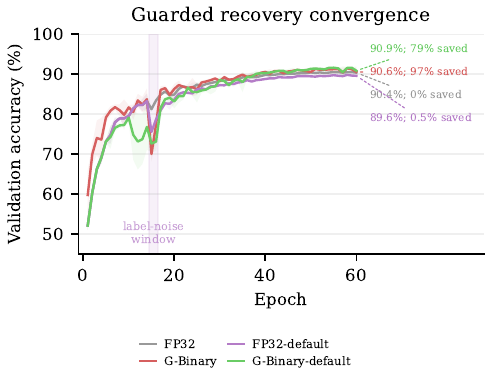}
  \caption{Guarded recovery pilot over 60 epochs.  Endpoint callouts report
  final validation accuracy and average traffic saved relative to \fp{}
  communication.}
    \Description{Control trace showing fallback to FP32 during an injected degradation window and re-admission after recovery.}
\label{fig:guarded-recovery}
\end{figure}
\endgroup

Beyond the recovery trace, the scoped pilots exercise the same interface in two
additional settings. On the CIFAR-100 live loop, the policy selects
\gter{}-backbone/\fp{}-head, matching the layer-aware boundary result. On
SST-2/DistilBERT, calibration transfers after tuning, with the all-\gbin{} policy reaches 90.71\%. Together, these pilots show that the controller interface can express warm-up,
admission, selective fallback, and recovery across more than one workload.

\section{Baseline Positioning}
\label{sec:positioning}

\sys{} occupies a different point from software compression, in-network
aggregation, and optimizer-only sign methods.  CXL pooling reduces
memory-placement pressure but still returns full-gradient traffic.  NCCL/RDMA
ring All-Reduce preserves \fp{} semantics but moves large gradient volumes.
Scalable Hierarchical Aggregation Protocol (SHARP)~\cite{sharp2016} and SwitchML-style aggregation ~\cite{switchml2021}
reduce network traffic, but they depend on switch support and do not place
aggregation in the CXL memory controller.
The two sign-gradient baselines serve different roles: MajoritySignSGD tests a
sign-only communication pattern, while SignOfMean tests the accuracy of taking a
sign after the \fp{} mean has already been computed.  \sys{} reduces memory-side gradient traffic when low-bit admission is
safe and keeps \fp{} as the calibration/recovery path when it is not.

The distinction is where the reduction happens and what information is
available when it happens.  MajoritySignSGD is communication-comparable to a
sign path because each worker can send a sign before aggregation, but its update
rule is still a software training method.  SignOfMean is an accuracy-oriented
reference: it first forms the \fp{} mean and then takes a sign, so it is not a
communication-equivalent controller primitive.  \sys{} moves this collective point to the CXL memory-controller boundary: admitted
gradient buckets are encoded as low-bit payloads before the write, and the
controller returns the selected aggregate on the read-response path while the
\fp{} software collective remains available for calibration and recovery.

\begin{figure*}[!t]
  \centering
  \includegraphics[width=0.9\textwidth]{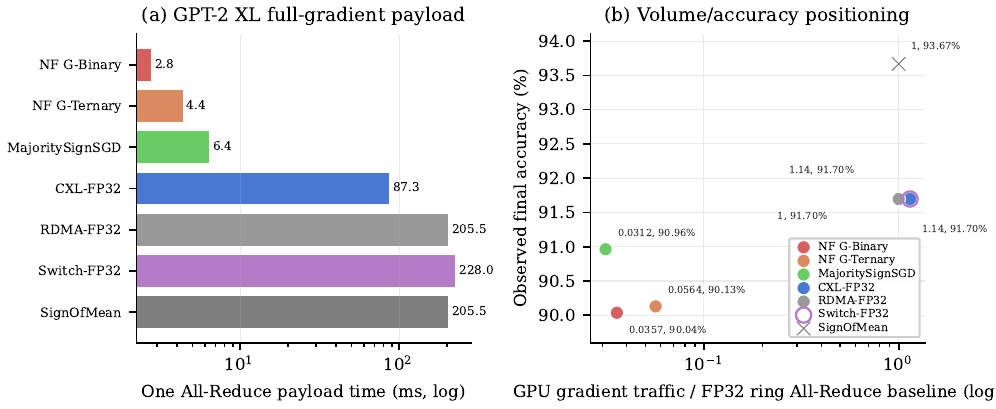}
  \caption{Modeled gradient-communication component for one GPT-2 XL payload.
  \sys{} occupies a controller-side low-bit point: lower modeled gradient
  traffic than \fp{} collectives when low-bit admission is valid, but not a
  universal replacement for software optimizers or switch-supported
  aggregation.}
    \Description{Baseline positioning plot comparing modeled communication time, traffic, and accuracy.}
\label{fig:baseline-position}
\end{figure*}

Fig.~\ref{fig:baseline-position} folds the traffic matrix into a two-panel view
of the GPT-2 XL payload.  Panel~(a) reports modeled communication time for one All-Reduce operation, and
panel~(b) positions the same paths by normalized traffic and observed
convergence evidence.  These numbers compare only the modeled
gradient-communication component under the stated payload and path assumptions;
they are not end-to-end training-step speedups.

The modeled communication time for \sys{} comes from two assumptions that
must both hold: gradient traffic reaches the CXL controller, and the admitted
workload can use packed \gbin{} or \gter{} aggregation.  In this model, the
controller-side \gbin{} and \gter{} rows take 2.77\,ms and 4.38\,ms for the GPT-2 XL
payload, compared with 87.31--228.00\,ms for the \fp{} collective references.
MajoritySignSGD has slightly lower normalized traffic than the \gbin{}
row, but it is not a controller-resident primitive.  SignOfMean is included
only to separate optimizer strength from communication equivalence, because it
requires the full-precision mean before taking the sign.

\section{Hardware Plausibility}
\label{sec:hardware}

\begin{figure}[!t]
  \centering
  \includegraphics[width=\linewidth]{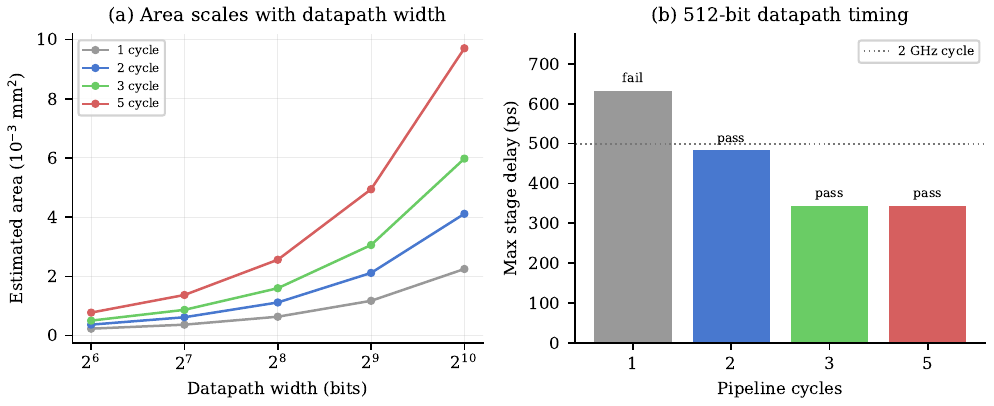}
  \caption{Hardware-cost model.  Area scales regularly with datapath width and
  pipeline depth; in the gate-equivalent model, the 512-bit five-cycle datapath has
  positive estimated slack against a 2\,GHz target.  The 500\,ps pass threshold
  in the timing panel is the cycle period of a 2\,GHz datapath stage.}
    \Description{Hardware-cost plot showing datapath-width area scaling and pipeline timing for the aggregation datapath.}
\label{fig:hardware-cost}
\end{figure}

The final question is whether the low-bit aggregation datapath is credible as
controller datapath logic, rather than an accelerator-scale block hidden behind
a timing constant.  We check this with three proxies: a gate-equivalent estimate
for the 512-bit five-cycle datapath, generic-cell synthesis over several
datapath widths, and an out-of-context FPGA place-and-route proxy.  The 512-bit point has positive
estimated slack against a 2\,GHz target, passes generic-cell synthesis, and
meets a 100\,MHz out-of-context FPGA route target.  Table~\ref{tab:hardware}
summarizes the corresponding area, energy, cell-count, and route checks.  These
proxy results do not replace foundry synthesis; they support treating the
aggregation datapath as a small regular datapath candidate rather than only a
timing placeholder.
For the timing sweep in Fig.~\ref{fig:hardware-cost}, the pass threshold is the
2\,GHz cycle period: 500\,ps.  A pipeline configuration therefore passes only
when its maximum stage delay fits within one 2\,GHz controller cycle.

The 512-bit datapath is one representative point in a width sweep.  The Yosys/ABC
generic-cell sweep~\cite{wolf2013yosys,abc} passes
structural checks for 64, 128, 256, 512, and 1024-bit datapaths, with total cells
growing from 606 at 64 bits to 8,646 at 1024 bits.  The nextpnr/Project Trellis ECP5 routing
proxy~\cite{nextpnr,projecttrellis} also passes across the same width range:
the 512-bit datapath routes at 184.88\,MHz and the 1024-bit stress point still
routes at 155.74\,MHz against a 100\,MHz target.  These checks do not prove a
2\,GHz ASIC implementation, but they provide structural support for treating
the five-cycle datapath as a plausible controller block.
Fig.~\ref{fig:hardware-cost} shows the scaling trend behind the 512-bit summary.

\begin{table}[!t]
  \centering
  \caption{Hardware plausibility summary for the 512-bit datapath.}
  \label{tab:hardware}
  \tablebodysetup
  \begin{tabular}{@{}lr@{}}
    \toprule
    Metric & Value \\
    \midrule
    Gate-equivalent estimate & 24,716 \\
    Estimated area & 0.004943\,mm$^2$ \\
    Estimated dynamic energy & 6.179\,pJ/line \\
    Estimated 2\,GHz slack & 155\,ps \\
    Yosys generic cells & 4,358 \\
    Yosys registers & 1,803 \\
    ECP5 routed target & 100\,MHz met \\
    \bottomrule
  \end{tabular}
\end{table}

\section{Limitations and Future Work}
\label{sec:limitations}

The evaluation assumptions in Section~\ref{sec:background} bound what this paper
claims: it evaluates a CXL controller primitive and an admission mechanism, not a
deployed end-to-end training runtime. The remaining limitations are therefore
policy and integration gaps around that primitive.

The policy results are also intentionally conservative.  The deterministic
ladder uses cosine-alignment thresholds from warm-up diagnostics, and the
Transformer pilot shows that thresholds calibrated on vision do not transfer
unchanged to fine-tuning.  The current paper therefore claims a controller
mechanism plus a measured admission boundary, not a universal policy oracle.
The correct next step is a workload-calibrated adaptive Commander/Supervisor
policy. Unlike the deterministic ladder used here, such a policy could combine
the signals already exposed by the prototype--layer diagnostics, loss trend, and
communication pressure--with richer training signals such as gradient norms and
update-history features.

The remaining production gap is metadata and runtime plumbing between the
training framework and the CXL-resident gradient buffers.  A complete system would need compiler or framework support to label parameter
groups and map those groups to CXL-resident gradient buffers. The distributed
training runtime would then attach mode metadata to collective operations,
encode packed sign or ternary payloads for admitted buckets, and account for
that encoding cost. Finally, recovery events must be reported back to the
training runtime so that fallback and re-admission remain visible to the
framework.
This is future integration work rather than an evaluation assumption: the
experiments evaluate whether the controller primitive and admission mechanism
are useful once such metadata and payload-encoding hooks are available.

\section{Conclusion}
\label{sec:conclusion}

Our simulations show that CXL-side low-bit gradient aggregation is a plausible
near-memory mechanism for distributed training, but only as a scoped
controller-level substrate.  The timing results show that a five-cycle
\gbin{}/\gter{} datapath has low exposed cost and can be hidden under bandwidth
pressure.  Functional tests show exact packed-sign semantics.  Training
evidence shows that low-bit aggregation is accurate on selected workloads, but
harder-workload boundary evidence rules out unconditional full-path low-bit
execution.
Layer-aware admission converts that boundary into a useful operating point:
low-bit backbone aggregation with an \fp{} classifier head preserves most of
the accuracy while reducing gradient traffic by more than 94\%.  The resulting
design principle is that \fp{} anchors calibration and recovery, while
\gbin{}/\gter{} provides a controller-side relief mode only when the workload,
layer, or phase admits it.

\begin{acks}
This work used computational resources provided by the Digital Research
Alliance of Canada, including access to the Narval cluster.
\end{acks}

\bibliographystyle{ACM-Reference-Format}
\bibliography{references}

\end{document}